\documentstyle[aps,prd,epsf]{revtex}

\renewcommand{\epsilon}{\varepsilon}
\begin{document}
\preprint{}
\title{Cosmic anti-friction and accelerated expansion}
\author{Winfried Zimdahl\footnote{Electronic address:
winfried.zimdahl@uni-konstanz.de}}
\address{Fachbereich Physik, Universit\"at Konstanz, PF M678
D-78457 Konstanz, Germany}
\author{Dominik J. Schwarz
\footnote{Electronic address: dschwarz@itp.hep.tuwien.ac.at}}
\address{Institut f\"ur Theoretische Physik, Technische Universit\"at Wien,
Wiedner Hauptstra\ss e 3--10/136,
A-1040 Wien, Austria}
\author{Alexander B. Balakin
\footnote{Electronic address: dulkyn@mail.ru}}
\address{Fachbereich Physik, Universit\"at Konstanz, PF  M678
D-78457 Konstanz, Germany\\
and
Department of General Relativity and Gravitation,
Kazan State University, 420008 Kazan, Russia\thanks{Present address}}
\author{Diego Pav\'{o}n\footnote{Electronic address:
diego@ulises.uab.es}}
\address{Departamento de F\'{\i}sica
Universidad Aut\'{o}noma de Barcelona,
08193 Bellaterra (Barcelona), Spain}

\date{\today}
\maketitle
\begin{abstract}
We explain an accelerated expansion of the present universe,
suggested from observations of supernovae of type Ia at high redshift, by
introducing an anti-frictional force that is self-consistently exerted on the
particles of the cosmic substratum.
Cosmic anti-friction, which is intimately
related to ``particle production'', is shown to  give rise
to an effective negative pressure of the cosmic medium.
While other explanations for an
accelerated expansion (cosmological constant, quintessence) introduce
a component of dark energy besides ``standard'' cold dark matter (CDM) we  
resort to a phenomenological one-component model of  CDM
with internal self-interactions.
We demonstrate how the dynamics of the $\Lambda$CDM model may be
recovered as a special case of cosmic anti-friction.
We discuss the connection with two-component models and obtain an attractor  
behavior for the ratio of the energy densities of both components which  
provides  a possible phenomenological solution to the coincidence problem.  
\end{abstract}
\vspace{1.5cm}

\section{Introduction}

There is evidence from SN Ia data for our present universe to be in a
state of accelerated expansion \cite{Riess,Schmidt,Perl,Zehavi,Riess2}.
This interpretation, which is indirectly also backed up by
recent data
from the  balloon experiments
Boomerang \cite{BOOM} and Maxima \cite{MAX} according to which
we live in a flat universe, requires a cosmic medium with
sufficiently high negative pressure to violate the strong energy
condition (SEC), $\rho + 3P > 0$. Cosmic matter with negative pressure is
now known as ``dark energy''. The problem why the density of the dark energy 
is of the order of the matter density just at the present epoch is the
``coincidence problem'' \cite{CDS}. An obvious dark energy candidate is a
cosmological constant of the order of the current critical density. Another
option is a scalar field called ``quintessence'', either with a
suitable potential  
\cite{Wette,Ratra,CDS,CCQ,PeeVi,Zlatev,Steinh,Martin,Barreiro,Luca}
or with a nonstandard kinetic term \cite{Armendariz}.

Negative pressures also occur in a different context. As was
first pointed out by Zel'dovich \cite{Zel} and Hu \cite{Hu},
quantum processes in the early universe such as cosmological
particle production may phenomenologically be equivalent to
effective negative bulk pressures. Numerous investigations have
subsequently explored this analogy
\cite{Turok,Barr,Calv,LiGer,ZP1,ZP2,ZP3,GaLeDe,Lima}. For
gaseous matter with specific internal self-interactions negative cosmic
bulk pressures can be derived within the framework of relativistic
gas dynamics \cite{ZTP,Zgeneq,ZiBa1,ZiBa2}. In particular, it
turned out that accelerated expansion (inflation) of the universe
can be driven by a self-interacting gas \cite{ZTP,Zgeneq,ZiBa1,ZiBa2}.

The observational data of the early 90th left ample room for a presently
non-negligible dissipative bulk stress on cosmological scales, as has been
pointed out by two of us in Ref.~\cite{PZ}. Hypothetically, this bulk stress 
was ascribed to internal interactions inside the dark matter, the latter being
supposed the dynamically dominating component of the universe. According to
the state of knowledge at that time our studies  were restricted to
matter for which the SEC holds. Taking into account the recent development
sketched above, it seems natural to investigate a present phase of
accelerated expansion from a  gas dynamical point of view.

The present work is based on the assumption that the observational evidence  
for an effective cosmological constant is an indication for the  existence  
of additional interactions  within the cosmic medium, which macroscopically
manifest themselves as negative pressures. Our strategy is the following:
Instead of introducing from the outset a new kind of matter (``quintessence'' 
or ``Q matter'') with a negative pressure in addition to cold  dark matter
(CDM), we start with a one-component description of the cosmic medium.
Although the corresponding substratum is assumed to consist of
non-relativistic particles, it does not represent simple dust since
we include  interactions within the matter. These interactions
turn out to be equivalent to effective one-particle forces, which are  
self-consistently exerted by the
cosmic medium on each of its individual particles. We show that the
cosmological principle restricts these forces to act like ``friction'' or
``anti-friction''.  Generalized equilibrium requirements for the cosmic medium
relate those forces to temperature and chemical potential. For a
non-relativistic substratum only cosmic anti-friction
generates a negative fluid bulk pressure.
We demonstrate that a suitable amount of
cosmic anti-friction leads to a SEC violation equivalent to an accelerated
expansion of the universe. On this basis we argue that the hypothesis of
cosmic anti-friction offers an alternative interpretation of the
magnitude-redshift relation for type Ia supernovae.

We present three different phenomenological models of cosmic
anti-friction which give rise to Hubble diagrams consistent with the
SN Ia data. We estimate the redshifts at which the accelerated expansion
started.
Although SN Ia data alone cannot discriminate between
those models, observations of the cosmic microwave radiation anisotropies
rule out one of them.

In a subsequent step we demonstrate how the anti-friction dynamics
may be decomposed into a two-component picture with one component being
Q matter, the other one  CDM. This procedure reveals that a negative
fluid pressure due to anti-friction  may be dynamically equivalent
to the accelerating effect of dominating Q matter. For a specific ansatz for 
the cosmic force and for a specific decomposition one recovers the
$\Lambda $CDM model. Moreover, it turns out that the same amount of
accelerated expansion of the presently observed universe is compatible with
different splittings of the total energy density into the energy densities of 
Q matter and CDM. In particular, there are decompositions in which Q matter
decays into CDM. For a specific decay rate there exists a stable attractor
solution with a fixed ratio of the energy densities of CDM and Q matter,
which indicates a possible solution of the coincidence problem.

The paper is organized as follows. Section II recalls some basic relations
and features on the possible role of a cosmic bulk pressure.
In Sec. III we we provide the basic relations of a kinetic theory for  
self-interacting gases.
We show that a certain class of particle number non-preserving interactions  
may be mapped onto effective one-particle forces.
The cosmological principle implies, that these forces necessarily describe a  
friction or an anti-friction within the cosmic medium.
We determine the strength of the force and realize
that only anti-friction is compatible with a negative pressure.
The back reaction of the anti-frictional self-interactions
on the cosmological dynamics is considered in Sec. IV for three different
anti-friction models. The luminosity and angular distances
as functions of redshift are studied for all three models.
Corresponding two-fluid models are established in Sec. V which
provide the basis for a discussion of the coincidence problem.
Sec. VI sums up our conclusions on the possible role of cosmic anti-friction.
Units have been chosen so that $c = k _{B} = \hbar = 1$.

\section{Cosmic bulk pressure}

We suppose the Universe to be describable by the stress-energy tensor of
an imperfect fluid
\begin{equation}
T^{ab}_{\rm eff} = (\rho + P) u^{a}u^{b} + P g^{ab} \ ,
\label{1}
\end{equation}
where $\rho $ is the energy density, measured by an observer comoving with
the fluid 4-velocity $u ^{a}$ normalized according to $u _{a}u ^{a}=-1$.
The effective pressure $P$ splits into two parts,
\begin{equation}
P = p + \Pi \ ,
\label{2}
\end{equation}
$p$ being the equilibrium pressure with $p\geq 0$ for gaseous matter
and $\Pi$ a non-equilibrium part. For a perfect fluid we have $\Pi = 0$,
i.e. $P\geq 0$. For a conventional viscous fluid $\Pi \leq 0$ is valid
during expansion, e.g., if kinetic energy of the fluid is transfered
to internal degrees of freedom. In the first-order Eckart theory
one has
$\Pi =-3H \zeta$ (see, e.g.\cite{Weinberg71}), where  $H \equiv u^a_{;a}/3 =  
\dot{a}/a$
is the Hubble expansion rate and  $\zeta \geq  0$ is the coefficient
of bulk viscosity. Within the more satisfactory second-order theories $\Pi$
becomes a dynamical degree of freedom (see, e.g., \cite{Roy,WZ,RM,Z2ndorder}).
Both the first- and the second-order theories are valid under the condition
$|\Pi|<p$, such that the effective pressure $P$  of a viscous fluid or gas
is positive.

Apart from viscosity, particle number non-conserving interactions inside
the matter may, as discussed above, lead to an effective bulk pressure.
This includes particle production out of the gravitational field.
The fact that $\Pi \leq 0$ if there is particle production and that $P<0$ is  
possible under such conditions may be demonstrated as follows.
Let the cosmic matter be characterized by the particle flow vector
\begin{equation}
N ^{i} = n u ^{i}\ ,
\label{3}
\end{equation}
where $n$ is the particle number density. In case the fluid particle number
is not preserved, the number density changes according to the balance
\cite{Calv,LiGer}
\begin{equation}
N^{a}_{;a}  =
\dot{n} + 3H n =  n \Gamma  \ ,
\label{4}
\end{equation}
where $\Gamma = \dot{N}/N$ is the change rate of the number
$N \equiv  n a ^{3}$  of particles in a comoving volume $a ^{3}$.
For $\Gamma > 0$ we have particle creation, for $\Gamma < 0$ particles are
annihilated. Conservation of the effective stress-energy tensor (\ref{1})
implies
\begin{equation}
\dot{\rho } + 3H \left(\rho + P \right) = 0 \ .
\label{5}
\end{equation}
With the help of the Gibbs equation
\begin{equation}
T \mbox{d}s = \mbox{d} \frac{\rho }{n} + p \mbox{d} \frac{1}{n}\ ,
\label{6}
\end{equation}
where $s$ is the entropy per particle and $T$ the temperature,
and  with  the balances (\ref{4}) and (\ref{5}) we obtain
\begin{equation}
nT \dot{s} = - 3H \Pi - \left(\rho + p \right)\Gamma \ .
\label{7}
\end{equation}
If the particle number $N$ is conserved, i.e. for $\Gamma =0$, the second
law of thermodynamics implies $\Pi \leq 0$
in an expanding universe.

If the particle number is not conserved one may define ``isentropic''
(or ``adiabatic'')
particle production by $\dot{s} = 0$,
which here means ``constant entropy per particle''.
Under this condition the equilibrium entropy
per particle does not change as it does in dissipative processes.
Instead, one can associate a viscous pressure to the particle production rate 
\cite{Calv,LiGer}:
\begin{equation}
\dot{s} = 0 \quad\Rightarrow\quad
\Pi = - \left(\rho + p \right)\frac{\Gamma }{3H }\ .
\label{8}
\end{equation}
The cosmic substratum is not a conventional dissipative fluid but a
perfect fluid with varying particle number.
Obviously, $\Gamma \geq 0$ guarantees $\Pi \leq 0$.
Substantial particle production is a phenomenon which is reasonably to
be expected in the early universe. It is less clear whether such processes
are operative at the present epoch as well. However, given that the nature
of CDM is unknown, there seems to be some room for speculations in this direction.
As easily seen, $P < 0$ in the case of dust ($p=0$) and $P \sim - \rho$
is possible if $\Gamma/(3H) = O(1)$. Compared to typical rates of particle
physics this requires an extremely small particle production rate only.
Below we shall comment on the origin of negative
bulk pressure in more detail.

Referring to matter creation as a relevant cosmological mechanism
may remind of corresponding processes within the steady state model \cite{Hoyle}. 
However, different from the latter, our considerations are fully within  
Einstein's theory.
Moreover, we shall trace the production process to internal interactions
within the system.

For a universe of the Friedmann-Lema\^{\i}tre-Robertson-Walker
(FLRW) type with scale factor $a$ we have
\begin{equation}
2 \frac{\ddot{a}}{a} + \left( \frac{\dot{a}}{a}\right)^{2} +
\frac{k}{a^{2}} = -8 \pi G P , \quad \quad (k = 1, 0, -1)
\label{9}
\end{equation}
and
\begin{equation}
\frac{k}{a^{2} H^{2}} = \Omega - 1, \qquad \Omega \equiv
\frac{\rho}{\rho_{c}} = \frac{8 \pi G}{3 H^{2}} \rho \ ,
\label{10}
\end{equation}
implying
\begin{equation}
\Pi  = - \frac{1}{4 \pi G} \left[\left(\frac{1}{2} \Omega - q \right) H^{2}
+ 4\pi G p \right]\ ,
\label{11}
\end{equation}
where $\, q \equiv - \ddot{a}/(a H^{2}) \,$ is the deceleration
parameter. In the standard big-bang scenario the non-equilibrium
pressure $\, \Pi \,$ is ignored and since $\, p \,$ cannot
become negative it folllows $\, \Omega (t) \leq 2 q(t)$.
But it is obvious that as long as $\, p(t) \,$
does not vanish and both pressures are of similar magnitude,
$\, \Omega (t) \,$ can be either larger or smaller than
$\, 2 q(t) \,$. Usually the current value of the hydrostatic
pressure is approximated by the state equation of dust
$\, p_{0} = 0$. The recent SN Ia data seem to confirm the existence
of an effectively negative pressure of the cosmic medium.
Assuming  $p_{0} = 0$  in Eq. (\ref{11}), the current value of the
bulk stress  can then be expressed as
\begin{equation}
\Pi_{0} = - \frac{1}{3} \left( 1 - 2
\frac{q_{0}}{\Omega_{0}} \right)\rho _{0}\ ,
\label{12}
\end{equation}
where the current energy density $\rho _{0}$ is of the order of the
critical energy density  $\rho _{c0}=3H_0^{2}/(8\pi G)$.
For $q _{0}>0$ we get $3 |\Pi_{0}|< \rho_{0}$, which is in
agreement with the strong energy condition.
The latter is violated, however, for $q _{0}<0$.
As already mentioned, for ``conventional'' viscous matter without
particle production the non-equilibrium part $\Pi$ of the
pressure is smaller in magnitude than the equilibrium contribution
$p$, so that $P>0$.
Here we argue that relation (\ref{8}) offers the option, to
understand the existence of a negative bulk pressure as a manifestation of
cosmological ``particle production''.
In the following sections we show how such a kind of negative pressure may
emerge as a consequence of specific internal interactions within the cosmic  
substratum.

\section{Kinetic theory for self-interacting gases}

\subsection{Basic relations}

The one-particle distribution function $f = f\left(x,p\right)$ of a
relativistic gas is supposed to obey the Boltzmann equation
\begin{equation}
L\left[f\right] \equiv
p^{i}f,_{i} - \Gamma^{i}_{kl}p^{k}p^{l}
\frac{\partial f}{\partial
p^{i}}
 = C\left[f\right] + {\cal S}\left(x, p\right)  \mbox{ , }
\label{13}
\end{equation}
where $C[f]$ is Boltzmann's collision integral.
The term ${\cal S}\left(x,p \right)$ on the right-hand side takes into  
account additional interactions which can not be reduced to elastic, binary  
interactions.
In particular, it describes production or decay processes of particles.

The particle number flow 4-vector
$N^{i}$ and the energy momentum tensor $T^{ik}$ are
defined in a standard way (see, e.g., \cite{Ehl}) as
\begin{equation}
N^{i} = \int \mbox{d}Pp^{i}f\left(x,p\right) \mbox{ , }
\ \ \
T^{ik} = \int \mbox{d}P p^{i}p^{k}f\left(x,p\right) \mbox{ .}
\label{14}
\end{equation}
The integrals in the definitions (\ref{14}) and in the following
are integrals over the entire mass shell, characterized by
$p^{i}p_{i} = - m^{2}$ and $p ^{0}>0$.
The entropy flow vector $S^{a}$ is given by \cite{Ehl}, \cite{IS}
\begin{equation}
S^{a} = - \int \mbox{d}P p^{a}\left[
f\ln f - f\right] \mbox{ , }
\label{15}
\end{equation}
where we have restricted ourselves to the case of
classical Maxwell-Boltzmann particles.

Using the general relationship \cite{Stew}
\begin{equation}
\left[\int p^{a_{1}}....p^{a_{n}}p^{b}f \mbox{d}P\right]_{;b}
= \int p^{a_{1}}...p^{a_{n}}L\left[f\right] \mbox{d}P
\label{16}
\end{equation}
and eq.(\ref{13}) we find
\begin{equation}
N^{a}_{;a} = \int \mbox{d}P\left(C\left[f\right] + {\cal S}\right)
\mbox{ , } \ \
T^{ak}_{\ ;k} =  \int \mbox{d}P p^{a}\left(C\left[f\right] + {\cal S}\right)
\mbox{ , }
\label{17}
\end{equation}
and
\begin{equation}
S ^{a}_{;a} \equiv  \sigma _{C} + \sigma _{{\cal S}} \ ,
\label{18}
\end{equation}
where
\begin{equation}
\sigma _{C} = - \int \mbox{d}P
C\left[f\right] \ln f
\mbox{ ,}
\label{19}
\end{equation}
and
\begin{equation}
\sigma _{{\cal S}} = - \int \mbox{d}P  {\cal S}\ln f
\mbox{ .}
\label{20}
\end{equation}

Under the condition  that
with respect to the elastic part of the interactions
the gas is at equilibrium, the
expression  $\ln f$
is a linear combination of the collision invariants
$1$ and $p^{a}$ and the collision integral $C \left[f \right]$
vanishes.
The corresponding equilibrium distribution function
becomes (see, e.g., \cite{Ehl})
\begin{equation}
f^{0}\left(x, p\right) =
\exp{\left[\alpha + \beta_{a}p^{a}\right] }
\mbox{ , }
\label{21}
\end{equation}
where $\alpha = \alpha\left(x\right)$ and
$\beta_{a}\left(x \right)$ is timelike.
This implies $\sigma _{C}=0$, i.e., there is entropy production only due to
$\sigma _{{\cal S}}$.

With $f$ replaced by $f^{0}$ in the definitions
(\ref{14}) and (\ref{15}), $N^{a}$, $T^{ab}$ and $S^{a}$ may be
split with respect to the unique 4-velocity $u^{a}$ according to
\begin{equation}
N^{a} = nu^{a} \mbox{ , \ \ }
T^{ab} = \rho u^{a}u^{b} + p h^{ab} \mbox{ , \ \ }
S^{a} = nsu^{a} \mbox{  . }
\label{22}
\end{equation}
Note that we have identified here the general fluid quantities $n$,
$\rho$ and $p$ of the previous sections with those emerging from the
Maxwell-Boltzmann gas dynamics. The exact integral expressions for $n$,
$\rho$, and $p$ may be found, e.g., in Ref. \cite{Groot}.
The entropy per particle $s$ is
\begin{equation}
s = \frac{\rho + p}{nT} - \frac{\mu }{T}\ .
\label{23}
\end{equation}
Here we have used the identifications $\beta _{i} = \beta u _{i}$,
$\beta = T ^{-1}$, and
$\alpha = \mu /T$  with $\mu $ being the chemical potential.

Use of the equilibrium distribution function (\ref{21}) in the
balances (\ref{17}) yields
\begin{equation}
\dot{n} + 3Hn
=   n \Gamma \equiv
\int \mbox{d}P {\cal S}^{0}\ ,
\label{24}
\end{equation}
and
\begin{equation}
u ^{a}\left[\dot{\rho }+3H \left(\rho + p \right) \right]
+ \left(\rho + p \right)\dot{u}^{a} + p _{,b}h ^{ab}
= - t ^{a}
\equiv
\int \mbox{d}P p ^{a}{\cal S}^{0}\ ,
\label{25}
\end{equation}
where ${\cal S}^{0}$ denotes the source term ${\cal S}$ for
$f=f ^{0}$.
We consider the special case that ${\cal S}^{0}$ depends linearly on
$f ^{0}$.
For reasons that will become clear shortly, we suppose that
the factor of proportionality can be written in terms of a suitable  
projection of a quantity $F ^{i}$ which will turn out to play the role of an  
effective one-particle force:
\begin{equation}
{\cal S}^{0} = -m \beta _{i}F ^{i}f ^{0}\ .
\label{26}
\end{equation}
The constant factor $-m$ has been chosen for later convenience.
The expression (\ref{26})  may be regarded as a special case of the more  
general structure
\begin{equation}
{\cal S} = -m F ^{i}\frac{\partial{f}}{\partial{p ^{i}}}\ .
\label{27}
\end{equation}
It is straightforward to realize that a ``collision'' term of this form may  
be taken to the left-hand side of Boltzmann's equation (\ref{13}),  resulting  
in
\begin{equation}
p^{i}f,_{i} - \Gamma^{i}_{kl}p^{k}p^{l}
\frac{\partial f}{\partial
p^{i}} + m F ^{i}\frac{\partial{f}}{\partial{p ^{i}}}
 = C\left[f\right]   \mbox{ . }
\label{28}
\end{equation}
The left-hand side of this equation can be regarded as
\[
\frac{\mbox{d}f \left(x,p \right)}{\mbox{d}\lambda }
\equiv  \frac{\partial{f}}{\partial{x ^{i}}}
\frac{\mbox{d}x ^{i}}{\mbox{d}\lambda }
+ \frac{\partial{f}}{\partial{p ^{i}}}
\frac{\mbox{d}p ^{i}}{\mbox{d}\lambda }
\]
with
\begin{equation}
\frac{\mbox{d} x ^{i}}{\mbox{d} \lambda  } = p ^{i}\ , \qquad
\frac{\mbox{D} p ^{i}}{\mbox{d} \lambda  } = m F ^{i}\ .
\label{29}
\end{equation}
Equations (\ref{29}) are the equations of motion for gas particles which  
move under the influence of a force field $F ^{i}=F ^{i}\left(x,p \right)$.
The quantity $\lambda  $ is a parameter along the particle worldline which for
massive particles may be related to the proper time $\tau $ by
$\lambda    = \tau /m$.
Consequently, a specific ``collisional'' interaction, described by a  
``source'' term ${\cal S}$, may be mapped onto an effective one-particle  
force $F ^{i}$.
This demonstrates that there exists a certain freedom to interpret  
collisional events in terms of forces. (This freedom can also been used in  
the reverse direction, i.e., to interprete (parts of forces) as collisions  
\cite{Kandrup}).
We emphasize that our approach is different from the ``canonical'' theory of  
particles in a force field for which the force term
$mF ^{i}\partial f/ \partial p ^{i}$  in Eq. (\ref{19}) is replaced by
$m\partial \left(F ^{i}f \right)/ \partial p ^{i}$ \cite{Kandrup}.
While both approaches are consistent with the equations of motion  
(\ref{29}), they coincide only for $\partial F ^{i}/ \partial p ^{i}=0$,  
which holds,
e.g., for the Lorentz force.
In the cases of interest here we will have
$\partial F ^{i}/ \partial p ^{i}\neq 0$.

\subsection{Cosmic forces}

Since we assume the universe to be homogeneous and isotropic at large scales 
we ask for forces which are consistent with the cosmological principle.
In such a case the metric of space-time is of the Robertson-Walker
form and the energy-momentum tensor is given by Eq. (\ref{1}).
Since the particle four-momenta are normalized
according to $p ^{i}p _{i} = - m ^{2}$, the force $F ^{i}$ has to
satisfy the relation $p _{i}F ^{i} = 0$. The momentum of a
comoving particle is $p^i _{\left(c \right)}= m u^i$. From its definition
comoving particles are force-free and thus $F^i(m u) = 0$.
This property follows also from the relation
\begin{equation}
\frac{\mbox{D} u ^{i}}{\mbox{d} \tau } = u ^{i}_{;n}\frac{p ^{n}}{m}\ ,
\label{30}
\end{equation}
which for $p ^{i}_{\left(c \right)}=mu ^{i}$ via
$u ^{i}_{;n}u ^{n}\equiv  \dot{u}^{i}\propto F ^{i}$ requires a vanishing
force since the cosmological principle implies $\dot{u}^{i}=0$.

On a spatial slice $\Sigma_t$, normal to $u^i$, the force field $F^i$ has to 
be independent of the spatial position, otherwise homogeneity would
be violated, thus $F^i = F^i(p;t)$. Since $p ^{i}$ and $u ^{i}$ are generally 
independent $4$-vectors, we may decompose according to
\begin{equation}
F^i = A p^i + B m u^i \ ,
\label{31}
\end{equation}
where $A$ and $B$ are arbitrary functions of $p^i$ with dimension $1/$time.
{}From $F^i p_i = 0$ we find
\begin{equation}
F^i = {B\over m} \left(- E p^i + m^2 u^i\right)\ ,
\label{32}
\end{equation}
with $E \equiv - p^i u_i$ being  the particle energy as measured by a
comoving observer. For a comoving particle one has $E=m$ and
we consistently recover that $F^i(m u) = 0$ for all $B$.
A particle which exactly moves with the mean macroscopic four-velocity is
force free. The temporal and spatial projections of the force field give
\begin{equation}
u_i F^i = {B \over m} (E^2 - m^2)\ , \qquad e_i F^i = - {B\over m}
E \sqrt{E^2 - m^2} \ ,
\label{33}
\end{equation}
where
\begin{equation}
e^i \equiv {1\over \sqrt{E^2 - m^2}} \left(p^i - E u^i\right) \
\label{34}
\end{equation}
is the spatial direction of the particle momentum
($e^i u_i = 0,\  e^i e_i = 1$).
Due to spatial isotropy $B$ may not depend on the spatial direction $e^i$,
thus $B = B(E; t)$. The expression (\ref{32}) is the most general force field
consistent with the cosmological principle.

According to the projections (\ref{33}), the force is acting parallel or
anti-parallel to the motion of the particle under consideration, depending
on the sign of $B$. For non-relativistic particles it should be
a good approximation to assume that $B$ is independent of $E$.
With $E = m + \epsilon $ where $\epsilon = m v ^{2}/2 \ll m$, we find at
leading order in the velocity $e_iF^i \simeq - B(m) m v$. This is nothing
but Stokes' law of friction. For $B > 0$ the force
field may be interpreted as cosmic friction, for $B<0$ as cosmic anti-friction. 
We conclude that cosmic (anti-)friction is the most general force field
which is compatible with the cosmological principle.

With the expressions (\ref{26}) for ${\cal S}^{0}$ and the equivalent force  
(\ref{32}) we may calculate the ``source'' terms in the balances (\ref{24})
and (\ref{25}).
For a $B$ independent of $E$,  the results are
\begin{equation}
\Gamma = - 3B\ ,
\label{35}
\end{equation}
and
\begin{equation}
 t ^{a}
= 3Bu ^{a}\left(\rho + p \right)\ .
\label{36}
\end{equation}
In general, neither the particle number nor the energy momentum are
conserved.
Consequently, $T ^{ab}$ in Eq. (\ref{22})  is {\it not} the quantity which  
will appear
on the right-hand side of Einsteins' field equations.
Eq. (\ref{35}) clarifies that the force strength determines the particle  
production (decay) rate
$\Gamma = \dot{N}/N$.
The entropy production density (\ref{18}) is determined by
\begin{equation}
S^{a}_{;a} = \sigma _{{\cal S}} =  - \alpha N^{a}_{;a}
- \beta_{a}T^{ab}_{\ ;b} = ns \Gamma = - 3nsB
\mbox{ . }
\label{37}
\end{equation}
A production of particles is characterized by $\Gamma >0$ and corresponds to  
$B<0$, i.e., an anti-frictional force, while a decay of particles is  
equivalent to an effective friction.

\subsection{Generalized equilibrium solutions}

To obtain the conditions under which the equilibrium distribution (\ref{21})
is preserved even
under the action of an (anti-)frictional force we insert
the expression (\ref{21}) into the Boltzmann equation (\ref{28}) which
yields
\begin{equation}
p^{a}\alpha_{,a} + \beta_{\left(a;b\right)}p^{a}p^{b}
= - m \beta _{i}F ^{i} \mbox{ .}
\label{38}
\end{equation}
Since $\beta _{i}\equiv  u _{i}/T$, it is only the projection
$u _{i}F ^{i}$ of the force which is relevant here.
If this projection vanishes,  relation (\ref{38}) reduces to the
``global'' equilibrium condition of standard relativistic kinetic
theory, i.e., to $\alpha = {\rm const}$ and either to the Killing vector
condition $\beta_{\left(a;b\right)}=0$ for $m > 0$, or to the conformal
Killing vector condition $\beta_{\left(a;b\right)}=\phi\left(x\right)g _{ab}$ 
for $m = 0$. The Friedmann models do not supply a timelike Killing vector,
thus there is no equilibrium solution for particles with finite mass.
However, in the non-relativistic limit $T\ll m$ a quasi-equilibrium,
characterized by $\alpha = m/T + {\rm const}$, $\beta_i = u_i/T$ and
$T \propto a^{-2}$ exists. In the case of cosmic (anti-)friction we find a
similar quasi-equilibrium solution for Friedmann models which reads
\begin{equation}
\label{nreq}
\alpha = \frac{m}{T}  + {\rm const}\ , \quad \beta_i = \frac{u _{i}}{T} \ ,
\quad
\frac{\dot{T}}{T} = - 2 \left(\frac{\dot{a}}{a} + B\right)  \ .
\label{39}
\end{equation}
With $3B = - \Gamma  = \dot{N}/N$ according to relation (\ref{35}), the
temperature behavior is
\begin{equation}
T \propto a ^{-2}N ^{2/3}\ .
\label{40}
\end{equation}
For vanishing (anti-)friction the particle number is constant and the familiar 
$T \propto a ^{-2}$ dependence for nonrelativistic matter is recovered.

The explicit knowledge of the force $F ^{i}$ in terms of $\Gamma $ now
allows us to study
the motion of the matter particles explicitly.
Contracting the equation of motion (\ref{29}) with the macroscopic
four-velocity results in
\begin{equation}
\frac{\mbox{D}\left(u _{i}p ^{i} \right)}{\mbox{d}\tau }
\equiv  - \frac{\mbox{d}E}{\mbox{d}\tau }
= u _{i}F ^{i} + \frac{1}{m}u _{i;k}p ^{i}p ^{k} \ .
\label{41}
\end{equation}
With Eq. (\ref{33}) and
under the condition of spatial homogeneity  we have
\begin{equation}
\frac{\mbox{d}E}{\mbox{d}\tau }
= - \frac{B+H}{m}\left(E ^{2} - m ^{2} \right)\ .
\label{42}
\end{equation}
Since $d \tau = d t \left(m/E \right)$ and $d E/d t \equiv
\dot{E} $ where $E = m + \epsilon $ with
$\epsilon \ll m$,
we finally obtain for the evolution of the non-relativistic (kinetic)
energy the expression,
\begin{equation}
\dot{\epsilon }
= - 2 \left(H - \frac{\Gamma }{3} \right)\epsilon \ .
\label{43}
\end{equation}
Consequently,
\begin{equation}
\epsilon \propto a ^{-2}N ^{2/3} \propto T \ ,
\label{44}
\end{equation}
the equipartition theorem.
Since the exponent $\alpha + \beta _{a}p ^{a} = \alpha -E/T $ of the
function (\ref{21}) in the nonrelativistic limit, with
$\alpha = m/T + {\rm const}$ and $E=m+\epsilon $, reduces to
$\alpha -E/T \rightarrow {\rm const}+ \epsilon /T$, relation (\ref{44})
demonstrates explicitly the invariance of the equilibrium distribution
(\ref{21}). The nonrelativistic velocity scales as
$v \propto a ^{-1}N ^{1/3}$.

{}With Eqs. (\ref{25}) and (\ref{36}) the fluid energy balance  becomes
\begin{equation}
\dot{\rho } + 3 H\left(\rho + p \right)
= - 3B \left(\rho + p \right) = \Gamma \left(\rho + p \right)\ .
\label{45}
\end{equation}
Introducing the quantity
\begin{equation}
\Pi \equiv   \frac{B}{H}\left(\rho + p \right)
= - \frac{\Gamma }{3H}\left(\rho + p \right)\ ,
\label{46}
\end{equation}
Eq. (\ref{45})  may be written in the form of the energy balance (\ref{5}),  
where $\Pi$ plays the role of an effective viscous pressure according to
the definition (\ref{2}), thus we have $P = p + \Pi$.
The energy balance for $T^{ab}$ with generally non-vanishing source in
(\ref{36}) is identical to the conservation law
$u _{a}T ^{ak}_{{\rm eff}\, ;k} = 0$  with
the energy momentum tensor (\ref{1}).
It is {\it this} energy-momentum tensor which appears on the
right-hand side of Einstein's field equations (see the discussion
following Eq. (\ref{36})).
With the re-interpretation (\ref{46}) of the sources on the right-hand side  
of the balance (\ref{25})  in terms of an effective pressure of the medium  
the latter becomes a ``closed'' system.
In the following we shall restrict ourselves to negative pressures $\Pi $,  
corresponding to the production of particles and to an anti-frictional force.  

We emphasize again that despite of the non-vanishing entropy production
$S ^{a}_{;a}$ the
microscopic particles are always governed by an equilibrium distribution
function.
In this context $S ^{a}_{;a}>0$ describes just an enlargement of the phase  
space of the system but not a dissipative process.
Although inter-particle collisions are necessary to establish an initial
equilibrium characterized by Eq. (\ref{21}), this equilibrium may then be
maintained even under the influence of the anti-frictional force and in the  
absence of further collisional interactions which might have been frozen out.  
The force which gives rise to a negative fluid pressure is
compatible  with an equilibrium distribution of the particles during the
expansion.
This feature is an essential advantage of the presented approach since it
allows us to apply standard gas dynamical concepts to characterize ``exotic''  
matter forms.

\section{Cosmological dynamics}

Let us assume the cosmic substratum after matter-radiation decoupling to be  
non-relativistic matter with
internal anti-friction, characterized by the energy-momentum tensor (\ref{1})  
with $p \ll \rho $ and
\begin{equation}
P \approx  \Pi = - \frac{\Gamma }{3H}\rho = - \frac{|B|}{H}\rho \ .
\label{47}
\end{equation}
Bearing in mind that $\Gamma = \dot{N}/N$ the energy
balance (\ref{45})
may be integrated to yield
\begin{equation}
\rho = \rho _{0}\frac{N}{N _{0}}\left(\frac{a _{0}}{a} \right)^{3}\ .
\label{48}
\end{equation}
The index $0$ again denotes the present epoch. According to the Friedmann equation 
for the spatially flat case, $8 \pi G\rho = 3 H ^{2}$, to which we restrict  
ourselves from now on,
the corresponding Hubble rate is given by
\begin{equation}
H = H _{0}
\left[\frac{N}{N _{0}}\left(\frac{a _{0}}{a} \right)^{3} \right]^{1/2}\ .
\label{49}
\end{equation}
For the ratio $\Pi / \rho $ we obtain
\begin{equation}
\frac{\Pi }{\rho } = \frac{B}{H}
= \frac{B}{H _{0}}
\left[\frac{N _{0}}{N }\left(\frac{a }{a _{0}} \right)^{3} \right]^{1/2}\ .
\label{50}
\end{equation}
The ``particle number'' changes as
\begin{equation}
N = N _{0}
\exp{\left[-3\int_{t _{}}^{t _{0}} \mbox{d}t|B| \right]}\ .
\label{51}
\end{equation}
For vanishing anti-friction, corresponding to a conserved particle number,
we have $N = N _{0}$.

The ratio $|B|/H$ enters the Hubble law for small redshift  $z$
($d _{L}$ is the luminosity distance),
\begin{equation}
H _{0}d _{L} = z + \frac{1}{2}\left(1-q _{0} \right)z ^{2} + ....\ ,
\label{52}
\end{equation}
via the deceleration parameter $q$, which for $k=0$ and nonrelativistic
matter becomes
\begin{equation}
q = \frac{1}{2} - \frac{3}{2}\frac{|B|}{H}\ .
\label{53}
\end{equation}
Generally, the luminosity distance $d _{L}$ in a spatially flat universe
may be written as
\begin{equation}
d_{\rm L} = \left(1+z\right)\int_{0}^{z} \frac{\mbox{d}z}{H \left(z \right)}\ . 
\label{54}
\end{equation}
As usual \cite{Weinberg} this is related to the angular distance $d_{\rm A}$  
of an object as
\begin{equation}
d_{\rm A}(z) = (1+z)^{-2} d_{\rm L}(z) \ .
\label{55}
\end{equation}
Relation (\ref{47}), the field equations
(\ref{9}) and (\ref{10}) for $k=0$ and $p \ll \rho $ may be combined to
yield
\begin{equation}
\frac{\Pi}{\rho} = \frac{B}{H} = - 1 - \frac{2}{3} \frac{\dot{H}}{H ^{2}}\ .
\label{56}
\end{equation}
It is convenient to express the Hubble rate as a function of redshift
$z = \left(a_0/a \right) - 1$.
With
\[
\dot{H} = - H ^{\prime }H (1 + z) \ ,
\]
where $H ^{\prime } \equiv  \mbox{d}H/ \mbox{d}z$,
the resulting equation is
\begin{equation}
\frac{H ^{\prime }}{B+H}
= \frac{3}{2(1+z)} \ .
\label{57}
\end{equation}
To establish specific models, assumptions about $B$ are necessary.
In the following we solve the cosmological dynamics for three different
choices of $B$ and compare the results with the Hubble diagrams from recent
SN Ia measurements. Although SN Ia data cannot discriminate between our
different models, since they provide equally good fits for redshifts
that are accessible with SN of type Ia, the additional information from
the angular distance to the last scattering surface and CMB measurements
will rule out one of the choices even without any detailed statistical analysis.

\subsection{The case $B \propto - H$}

\begin{figure}
\begin{center}
\epsfig{figure=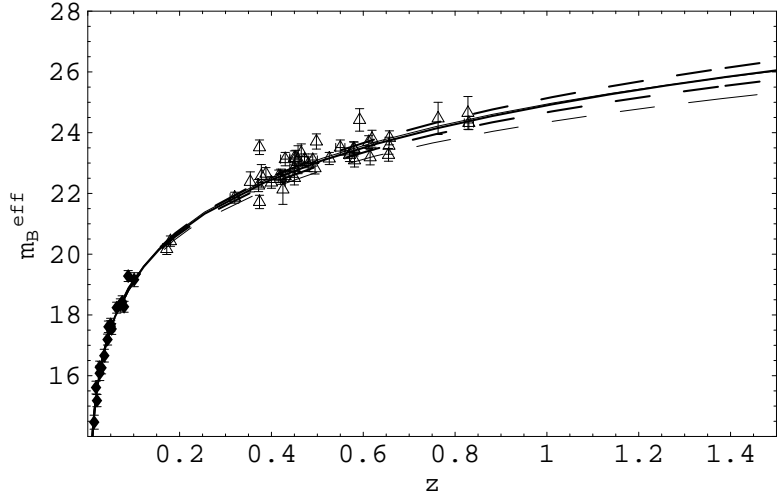,width=0.6\linewidth}
\end{center}
\caption{Hubble diagram for  $(\Omega_M, \Omega_\Lambda) = (1,0)$ [thin, dashed 
line], $(0.3,0.7)$ [thin line] compared to case A with the values $\nu = 
0.7,0.5,0.3$ [thick lines from top to bottom]. The data points are taken from 
Perlmutter et al. 1999, the diamonds are the Cal\'an/Tolodo SN1a data, the 
triangles are those of the SCP.}
\label{fig1}
\end{figure}

\begin{figure}
\begin{center}
\epsfig{figure=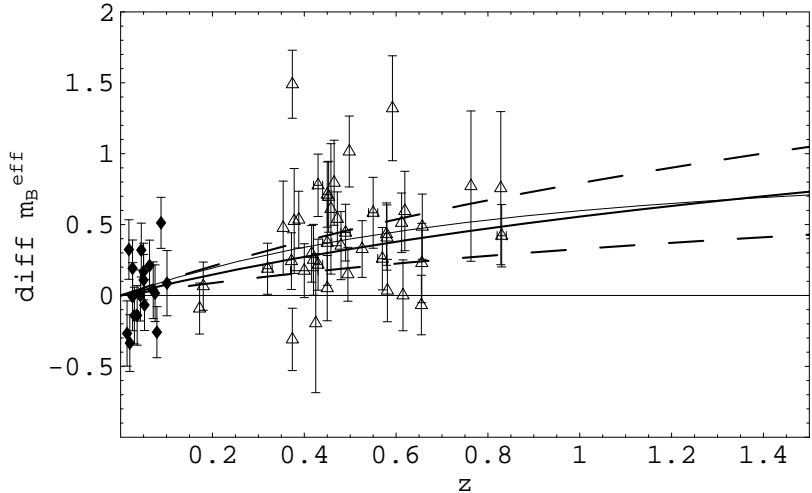,width=0.6\linewidth}
\end{center}
\caption{Differences of the magnitudes with respect to a
$(\Omega_M, \Omega_\Lambda) = (1,0)$ universe versus redshift. The plotted
models and data are the same as in figure 1. Note that other authors often
depict the magnitude difference with respect to an empty universe, which
has negative spatial curvature and is thus incompatible with inflation.}
\label{fig2}
\end{figure}

The simplest possible choice is apparently $B = -\nu H$ with $\nu ={\rm const}$.
According to Eq. (\ref{47}) this ansatz is equivalent to assuming a constant  
ratio $\Pi/\rho$,  which should be considered realistic at most piecewise.
With this assumption we get a Hubble rate
\begin{equation}
H(z) = H _{0}(1 + z)^{\frac{3}{2}(1-\nu )}\ .
\label{58}
\end{equation}
Note that the same power-law behavior follows if a perfect fluid with
the equation of state $- \nu = w = p/\rho$ is assumed. We recover the matter
dominated universe for $\nu = 0$ and the vacuum dominated universe
for $\nu = 1$. Equation (\ref{54})
is easily integrated to provide an explicit expression for the luminosity
distance ($\nu \neq 1/3$)
\begin{equation}
H _{0}d _{L} = \frac{2}{3 \nu - 1}
\left[\left(1+z \right)^{\frac{1}{2}\left(1+3 \nu  \right)} - 1 - z \right]\ .
\label{59}
\end{equation}
This result coincides with the one by Lima and Alcaniz \cite{LimaAlc}.
The corresponding luminosity distance - redshift relation is
shown in Fig.~\ref{fig1} while
Fig.~\ref{fig2} shows the difference to the Standard Cold Dark Matter (SCDM)  
model.
We find that $\nu = 0.5$ gives a good fit to
the observations
of SN Ia.
At first sight one might have expected the best fit for $\nu = 0.7$
since it is this value which reproduces an equation of state $P/\rho =
P_{0}/\rho_{0}=-0.7$. However, as already indicated, a constant ratio
$P/ \rho $ is realistic only piecewise. The fact that the better fit is
$\nu = 0.5$ suggests that $\nu $ must have been smaller for $z > 0$ than
for $z=0$.
This illustrates the circumstance that the SN Ia observations for higher $z$  
do not directly reflect the cosmological equation of state at the present
time but, as to be seen from the integral in the expression (\ref{54}) for
$d _{L}$, depend on the entire dynamics from redshift $z$ to redshift
$0$.

The ``particle number'' changes according to
\begin{equation}
\frac{N}{N_0} =
\exp\left(-3\int_0^z\frac{|B|}{H}{{\rm d}z\over(1 + z)}\right) =
(1+z)^{- 3 \nu} \ .
\label{60}
\end{equation}
The energy density [cf. Eq. (\ref{48})] decays as $\rho \propto
(1+z)^{3(1-\nu)}$, the temperature [cf. Eq. (\ref{40})] as $T
\propto (1+z)^{2(1-\nu)}$. For the 'best fit' $\nu = 0.5$ there is a
decrease in $N/N_0$ by about 3(5) for a redshift of 1(2).
Over a larger interval $N$ might decrease by many orders of magnitude.
This again reflects the unrealistic nature of $P/ \rho ={\rm const}$ over a  
large range of $z$ values.

The observations of SN type Ia are restricted to redshifts of order 1.
In order to test a specific model of anti-friction at redshift $z \gg 1$,
additional information is required.
A very promising possibility are CMB anisotropy data.
Acoustic oscillations in the early
universe give rise to peaks and dips in the band power spectrum of these
anisotropies.
The typical length scale of the
largest acoustic oscillations is given by the sound horizon $R_{\rm s} =
c_{\rm s}/H$ at the time of photon decoupling, where $c_{\rm s}$ is the  
sound velocity.
This physical length corresponds to the presently
observed angular scale \cite{Weinberg} of the first acoustic peak
\begin{equation}
\delta_{\rm 1st\ peak} \approx {R_{\rm s}(z_{\rm dec})\over
d_{\rm A}(z_{\rm dec})} \ .
\label{61}
\end{equation}
In fact, the corresponding observational data  rule out any constant value  
for $P/\rho$.  Namely,
insertion of Eqs.~(\ref{58}), (\ref{55}) and (\ref{59}) leads to ($z_{\rm dec}
\gg 1$)
\begin{equation}
\delta_{\rm 1st\ peak} \approx
c_{\rm s} \frac{3\nu - 1}{2} \approx 0.14 {\rm rad}\ ,
\label{62}
\end{equation}
which for $\nu =0.5$ and $c _{s}=\sqrt{1/3}$  exceeds the detected  
$\delta_{\rm 1st\ peak} \simeq 0.9^\circ
= 0.016$ rad \cite{BOOM,MAX} by about one order of magnitude.
Thus, a constant effective equation
of state over a large range in redshift is incompatible with observations.
To illustrate this feature we plot in Fig.~\ref{fig7} the angular scale  
under which the Hubble radius $H ^{-1}\left(z \right)$ is seen for the
`best fit' $\nu =0.5$.
At
$z=z _{{\rm dec}} \approx 1100$ the Hubble radius differs from the sound  
horizon only by a factor $c _{s}\approx \sqrt{1/3}$.
The present model corresponds to the upper curve.
In the following we consider two cases for which $P/\rho$ is time varying.
Since we are interested in a modification of the cosmological dynamics at
late times we focus on two models where this ratio is increasing with time.  

\subsection{The case $B = - |B _{0}| = {\rm const}$ }

\begin{figure}
\begin{center}
\epsfig{figure=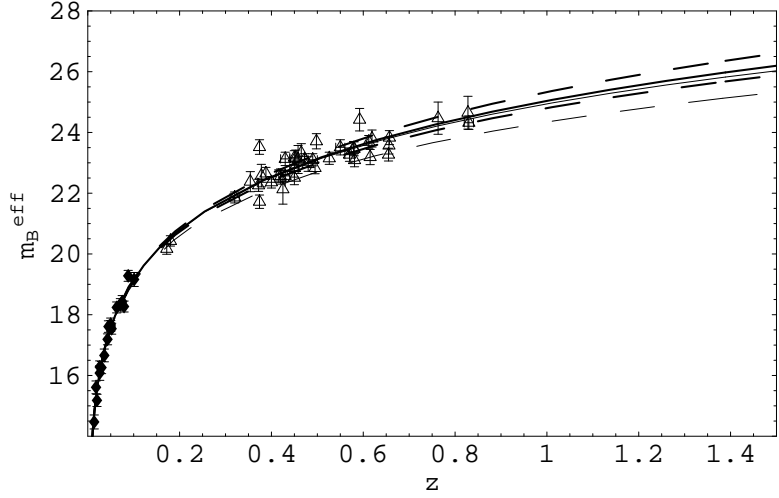,width=0.6\linewidth}
\end{center}
\caption{Same as in figure 1 but for case B with $|B_0|/H_0 =
0.9,0.7,0.5$ [thick lines from top to bottom].} 
\label{fig3}
\end{figure}

\begin{figure}
\begin{center}
\epsfig{figure=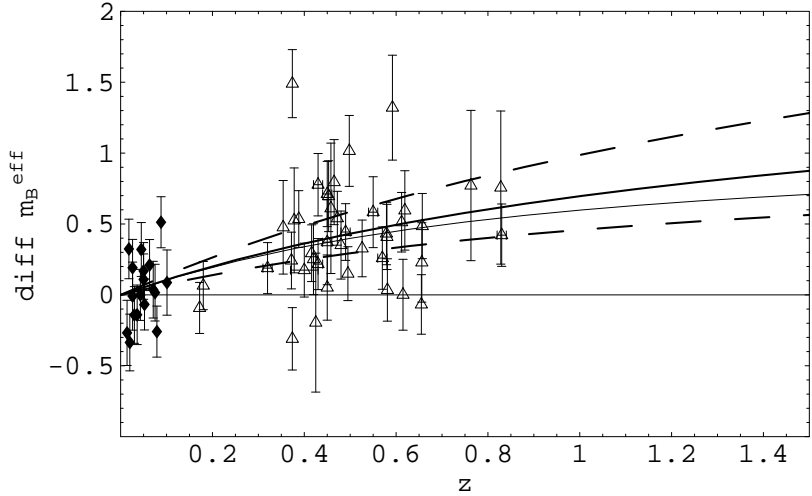,width=0.6\linewidth}
\end{center}
\caption{Differences of the magnitudes versus redshift for the
models in figure 3.}
\label{fig4}
\end{figure}

{}From Eq. (\ref{47}) follows that the simplest choice leading to an
increasing ratio $|\Pi|/\rho$ is a constant value of $B$.
{}From Eq. (\ref{57}) we obtain with $\nu_0 \equiv |B_0|/H_0$,
\begin{equation}
H(z) = H_0[(1 - \nu_0)(1+z)^{3/2} + \nu_0] \ .
\label{63}
\end{equation}
This solution describes a transition from matter domination
to vacuum domination of the universe. In the past, $z \gg 1$, we have
$H \propto (1+z)^{2/3}$, a matter dominated universe, while for the future,
$z \to -1$, $H$ and thus $\rho$ are constant. The expansion of the
universe starts to accelerate when $H(z) < 3 |B|$ [cf. Eq.(\ref{53})], which  
happens for
a redshift $z_{\rm acc}$, given by
\begin{equation}
1 + z_{\rm acc} = \left[{2 \nu_0 \over 1 - \nu_0}\right]^{2/3} \ .
\label{64}
\end{equation}
In Figs.~\ref{fig3} and \ref{fig4}  we compare the Hubble diagram for this  
model to recent
observations of SN Ia. Here we find a `best fit' value $\nu_0 = 0.7$.
For this model the universe starts acceleration at $z_{\rm acc} \approx 1.8$. 
The increase in the ``particle number'' may be expressed as
\begin{equation}
\frac{N}{N_0} = \frac{\left[(1 -\nu_0)(1+z)^{3/2} + \nu_0 \right]^{2}}
{\left(1+z \right)^{3}}\ ,
\label{65}
\end{equation}
which gives an increase by a factor of about $5$ since $z_{\rm acc}$.

Also for this model the angular scale of the Hubble radius is shown in  
Fig.~\ref{fig7} for $\nu _{0}=0.7$  (second curve).
Obviously, here the Hubble scale is of the order of the scale of the first  
acoustic peak, i.e., this model is consistent with the CMB observations.
However, a more quantitative statement could only be made
on the basis of a Boltzmann code, which is beyond the scope of our present
work.

\subsection{The case $B \propto - H ^{-1}$ }

\begin{figure}
\begin{center}
\epsfig{figure=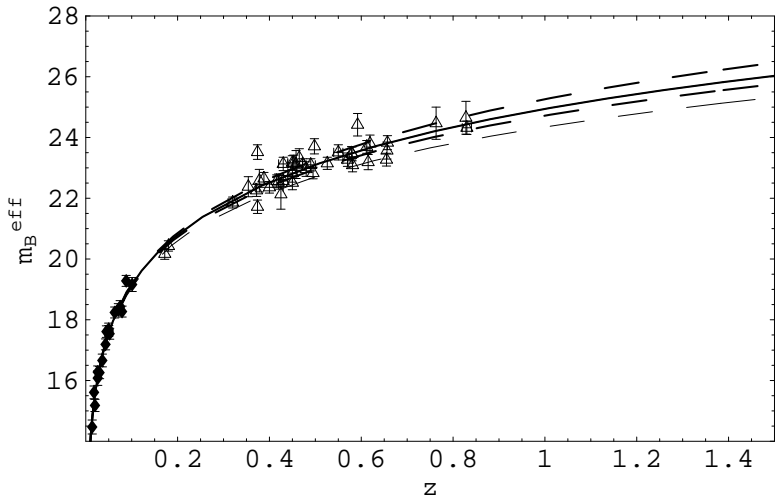,width=0.6\linewidth}
\end{center}
\caption{Same as in figure 1 but for case C with $1/(1+\mu) =
0.9,0.7,0.5$ [thick lines from top to bottom]. Note that this case is 
identical to $\Lambda$CDM, with $\Omega_\Lambda = 1/(1+\mu)$.}
\label{fig5}
\end{figure}

\begin{figure}
\begin{center}
\epsfig{figure=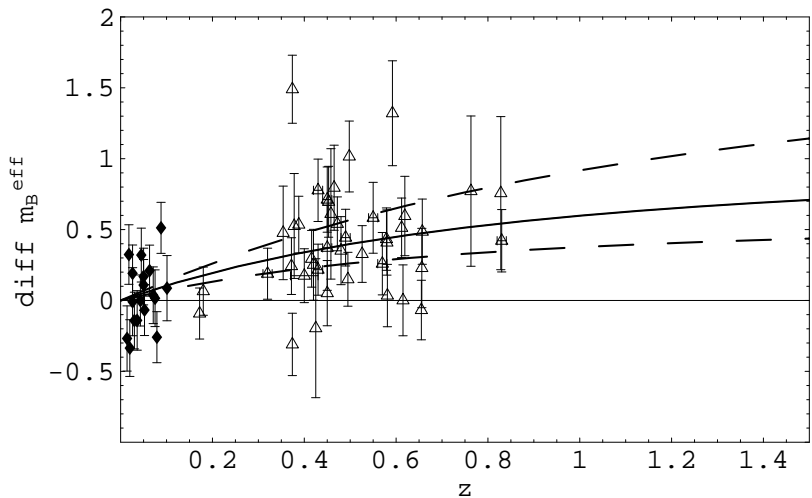,width=0.6\linewidth}
\end{center}
\caption{Differences of the magnitudes versus redshift for the models in  
figure 5.}
\label{fig6}
\end{figure}

Another choice leading to an increasing ratio $|\Pi| / \rho $
is $|B| \propto H ^{-1}$, equivalent to an ansatz
\begin{equation}
\frac{|B|}{H} = \frac{1}{\mu + 1}\frac{H _{0}^{2}}{H ^{2}}\ .
\label{66}
\end{equation}
The specific choice of the constant factor
$1/ \left(\mu  +1 \right) $ was made for later convenience.
Integration of Eq. (\ref{57})  with the ansatz (\ref{66}) yields
\begin{equation}
H = {H_0\over \sqrt{1 + \mu}}\left[\mu (1+z)^3 + 1\right]^{1/2}\ .
\label{67}
\end{equation}
For $z \gg 1$ we have
\begin{equation}
H \propto (1+z)^{3/2} \ ,
\label{68}
\end{equation}
which is characteristic of a matter-dominated universe.
For the opposite case $z \to -1$ the Hubble rate
approaches the constant value
\begin{equation}
H \rightarrow \frac{H_0}{\sqrt{\mu + 1}} \ .
\label{69}
\end{equation}
The Hubble rate (\ref{67}) implies again a transition from a
matter-dominated universe at $z\gg 1$ to a de Sitter
universe as $z \to -1$. Equations (\ref{67}) and (\ref{54})
determine the luminosity distance, which is plotted in Figs.~\ref{fig5} and
\ref{fig6}. This
case in fact includes all $\Lambda$CDM models, as can be easily seen
by replacing $1/(\mu +1) \to \Omega_\Lambda$ and $\mu /(\mu +1) \to
\Omega_{\rm CDM}$. As expected, the 'best fit' model has $1/(1+\mu) = 0.7$.

To obtain the redshift at which acceleration starts we write
\begin{equation}
\frac{|B|}{H} = \frac{1}{\mu (1+z)^3 + 1} \ .
\label{70}
\end{equation}
The condition $|B|/H \geq 1/3$ for accelerated expansion [cf. Eq. (\ref{53})] 
gives $1 + z_{\rm acc} = (2/\mu)^{1/3}$. The growth of $N$ follows from
\begin{equation}
N = N_0\frac{\mu(1+z)^3 + 1}{(\mu + 1)(1+z)^3}\ .
\label{71}
\end{equation}
For the favored value $1/(1+\mu) = 0.7$ accelerated expansion starts at
$z_{\rm acc}\approx 0.67$, which is in the expected range
$0.5 < z _{\rm acc} < 1$ [cf. Ref. \cite{Riess01}], and $N$ has grown since then
by a factor of about $3$.

\begin{figure}
\begin{center}
\epsfig{figure=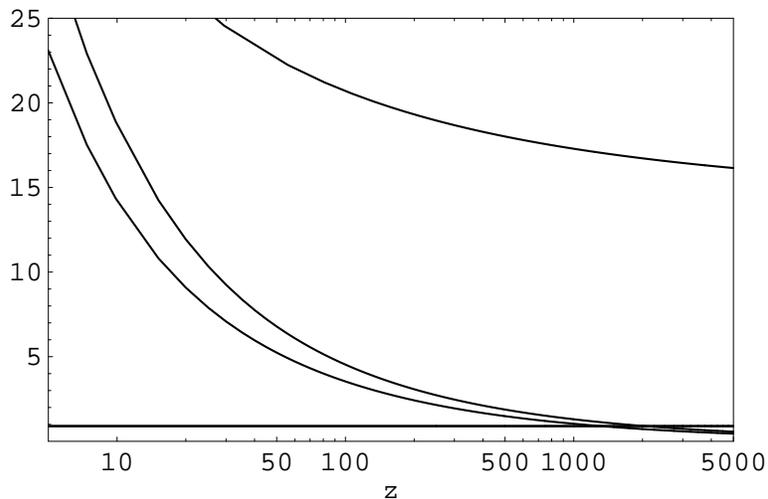,width=0.6\linewidth}
\end{center}
\caption{The angular scale
$H ^{-1}(z)/d_{\rm A}(z)$ in degrees under which the
Hubble radius $H ^{-1}$
is seen for the models of subsections ~IV.A., ~IV.B., and ~IV.C (from top to  
bottom).
For comparison we indicate
the angular scale of the observed first acoustic peak
($\simeq 0.9^\circ$)  by the horizontal line.}
\label{fig7}
\end{figure}

Again we show an estimate of the angular scale of the Hubble radius
(for $1/ \left(1+\mu \right)=0.7$) in comparison to the scale of the first  
acoustic peak in
Fig.~\ref{fig7} (lower curve). This model is consistent with
observations as well.
The difference to the former model is large
enough that it should give rise to a significant difference in the CMB
predictions. We do not further elaborate this point here.

\section{Two component models and the coincidence problem}

\subsection{Interacting fluids}

In this section we investigate in which sense cosmological anti-friction may  
be regarded as effective description for an underlying two-component model.
To this purpose we decompose the total energy density (\ref{48})
into a ``conventional'' matter part (subscript M) and a
``Q matter'' part (subscript Q):
\begin{equation}
\rho = \rho _{M} + \rho _{Q}\ ,
\label{72}
\end{equation}
with
\begin{equation}
\rho _{M} \equiv  r \rho _{0}\left(\frac{a _{0}}{a} \right)^{3}
\mbox{\ \ \ }\mbox{\ \ \ }\mbox{\ \ \ }\mbox{\ \ \ }
{\rm and}
\mbox{\ \ \ }\mbox{\ \ \ }\mbox{\ \ \ }\mbox{\ \ \ }
\rho _{Q} \equiv  \left[\frac{N}{N _{0}} - r \right]
\rho _{0}\left(\frac{a _{0}}{a} \right)^{3}\ ,
\label{73}
\end{equation}
where $r=r \left(a \right) > 0 $ with $r<N/N _{0}$ and
$r _{0}\equiv  r \left(a _{0} \right)<1$.
The factor $r _{0}$ fixes the ratio of both components at $a _{0}$:
\begin{equation}
\frac{\rho  _{M}\left(a _{0} \right)}{\rho _{Q}\left(a _{0} \right)}
= \frac{r _{0}}{1-r _{0}}\ .
\label{74}
\end{equation}
For $r = {\rm const}$ the universe evolves as though the cosmic
substratum would consist of two non-interacting components: Non-relativistic 
matter with $p _{M}=0$ and Q-matter with an equation of state
\begin{equation}
P _{Q}= \Pi
= \frac{B}{H}\rho = -\frac{|B|}{H}\frac{N/N _{0}}{(N/N _{0}) - r}
\rho _{Q}\ ,
\quad \left(r= {\rm const} \right)
 \ .
\label{75}
\end{equation}
For an expanding universe the pressure $\Pi$ is always negative.
For any form of the potential of a quintessence model the corresponding  
coefficient of anti-friction is easily obtained from Eq. (\ref{75}).

If $r$ is allowed to vary  we have
\begin{equation}
\dot{\rho }_{M} + 3 H \rho _{M} = \frac{\dot{r}}{r}\rho _{M}\ .
\label{76}
\end{equation}
For $\dot{r}>0$ the term on the right-hand side may be regarded as a matter
``source''. This corresponds to a ``sink'' in the energy balance of the Q
component,
\begin{equation}
\dot{\rho }_{Q} + 3 H \rho _{Q} = \Gamma \left(\rho _{Q} + \rho _{M} \right) 
- \frac{\dot{r}}{r}\rho _{M}\ .
\label{77}
\end{equation}
By introducing the effective pressures
\begin{equation}
P _{M} = \Pi _{M} \equiv  - \frac{\dot{r}/r}{3H}\rho _{M}\ ,
\label{78}
\end{equation}
and
\begin{equation}
P _{Q} =  \Pi - \Pi _{M}
= - \left[\frac{\Gamma }{3H}\frac{N}{N _{0}}
- \frac{\dot{r}}{3H}\right]
\frac{\rho _{Q}}{(N/N _{0})- r}\ ,
\label{79}
\end{equation}
the balances (\ref{76})  and (\ref{77})  become
\begin{equation}
\dot{\rho }_{M} + 3H \left(\rho _{M} + P _{M} \right) = 0
\label{80}
\end{equation}
and
\begin{equation}
\dot{\rho } _{Q} + 3H \left(\rho _{Q} + P _{Q} \right) = 0 \ ,
\label{81}
\end{equation}
respectively.
The circumstance that for $\dot{r}>0$ {\it both} components have a negative  
effective pressure (given that
$|\Pi _{M}|<|\Pi |$) is essential for a discussion of
the ``coincidence problem'' in the present context.

Supernovae Ia observations suggest that $|P|/ \rho = |B|/H$ is of the order  
unity in the present epoch.  Within our approach this raises the question:
Why is the coefficient of antifriction $-B$ of the order of the Hubble rate  
just at the present epoch?
As demonstrated in Sec. IV.A. for the case $B \propto -H$, the CMB data rule  
out that this was the case through the entire evolution of the universe  
since decoupling.
Thus it seems to be a coincidence that we live in this special era.
The decomposition (\ref{72}) allows us to relate our consideration to the  
usual discussion of the coincidence problem.
It is obvious that all models according to which the universe is made of a  
non-interacting mixture of quintessence and CDM correspond to
$r={\rm const}$. In this special case our approach does not provide any new  
insight into the coincidence problem.
For $\dot{r}>0$, however, it offers a solution which is similar to the one    
proposed by Chimento et al.
\cite{ChiJaPa}.
The quantity of interest is the ratio $\rho _{M}/ \rho _{Q}$ which is
governed by the equation
\begin{equation}
\left(\frac{\rho _{M}}{\rho _{Q}} \right)^{\displaystyle \cdot}
= 3H \left[\frac{\rho _{M}}{\rho _{Q}} \right]
\left[\frac{P _{Q}}{\rho _{Q}}
- \frac{P _{M}}{\rho _{M}}\right]\ .
\label{82}
\end{equation}
For $P _{M}=0$, or equivalently $r={\rm const}$, one has
$P _{Q}=\Pi <0$, i.e., the ratio $\rho _{M}/ \rho _{Q}$ continuously
decreases and for large cosmological times one has
$\rho _{M} \ll \rho _{Q}$. In other words, the matter component becomes
dynamically negligible.
However, if an exchange between both components is admitted, which amounts
to a nonvanishing quantity $P _{M}=\Pi _{M}$, there exists a second stationary
solution of Eq. (\ref{82}), namely
\begin{equation}
\frac{P _{Q}}{\rho _{Q}}
= \frac{P _{M}}{\rho _{M}}\ .
\label{83}
\end{equation}
{}Combining relations (\ref{78})  and (\ref{79})  we obtain
\begin{equation}
\frac{P _{Q}}{\rho _{Q}}
- \frac{P_{M}}{\rho _{M}}
= \frac{1}{3H}\frac{N/N _{0}}{(N/N _{0}) - r}
\left[\frac{\dot{r}}{r} -\Gamma \right]\ .
\label{84}
\end{equation}
Obviously, the condition  (\ref{83}) is equivalent to
\begin{equation}
\frac{\dot{r}}{r} = \frac{\dot{N}}{N}
\quad\Rightarrow\quad r = r _{0}\frac{N}{N _{0}} .
\label{85}
\end{equation}
Via Eq. (\ref{73}) the stationarity  condition (\ref{85})  for the
ratio $\rho _{M}/ \rho _{Q}$  provides us with
\begin{equation}
\left(\frac{\rho _{M}}{\rho _{Q}} \right)_{s} = \frac{r _{0}}{1 - r _{0}}\ ,
\label{86}
\end{equation}
which, according to the split (\ref{73}),  is just the present ratio of both
components.
Concerning its possible role for the coincidence problem
it is interesting to investigate the stability properties of the solution
(\ref{86}).
To this purpose we consider deviations from this stationary value:
\begin{equation}
\frac{\rho _{M}}{\rho _{Q}}
= \left(\frac{\rho _{M}}{\rho _{Q}} \right)_{s} + \delta \ .
\label{87}
\end{equation}
Since $\left(\rho _{M}/ \rho _{Q} \right)_{s}^{\displaystyle \cdot}=0$,
the resulting equation is
\begin{equation}
\dot{\delta }= 3H
\left[\left(\frac{\rho _{M}}{\rho _{Q}} \right)_{s} + \delta  \right]
\left[\frac{P _{Q}}{\rho _{Q}}
- \frac{P _{M}}{\rho _{M}}\right]\ .
\label{88}
\end{equation}
By visualizing this dynamics in a $\dot{\delta }- \delta$ diagram,
the stationary solution (\ref{86}) is an attractor solution for
\begin{equation}
\frac{P _{Q}}{\rho _{Q}}
- \frac{P _{M}}{\rho _{M}} < 0 \quad\Rightarrow\quad
0 < \frac{\dot{r}}{r} < \Gamma = \frac{\dot{N}}{N}\ .
\label{89}
\end{equation}
The rate $\dot{N}/N$ represents a limit for the rate $\dot{r}/r$ at which
energy is transfered from the Q component to the matter.
We conclude that a certain decay of vacuum like Q matter into CDM offers
a potentially promising approach to the coincidence problem.
The phenomenological concept of a decaying vacuum is widely used in the
physics of the early universe in order to describe the decreasing dynamical  
role of an effective cosmological ``constant'' connected with a transition
from an initial inflationary period to a subsequent FLRW behavior
\cite{Ozer,GaLeDeCQG}. It is note-worthy that a similar mechanism seems to
be relevant also for the transition from matter dominance to ``vacuum''
dominance in the late universe.
At first sight it might seem counter-intuitive that a decay of the vacuum at  
the same time leads to an apparent dominance of the latter. However,
for the previously discussed model
$B \propto -H ^{-1}$ (Sec.~IV.C.) we shall confirm explicitly that such
kind of transition is indeed consistent with a positive rate $\dot{r}/r>0$.

At this point we come back to our previous statements
(following Eqs. (\ref{8}) and (\ref{46})) on the nature of ``particle  
production'' in the
presented formalism.
According to Eq. (\ref{75}) the ratio $|B|/H$, equivalent to $\Gamma /3H$
determines the contribution of the ``vacuum'' to the total energy density.
Cosmic anti-friction may be viewed as a vacuum effect which is connected
with a
nonvanishing ``particle production'' rate $\Gamma $.
For an equation of state $\rho _{M}=n _{M}m$ it is obvious, that
$\dot{r}/r$ is the production rate of matter particles out of the
decaying ``vacuum'' component Q.
It represents
a real ``physical''  particle production.
According to Eq. (\ref{89}) the rate $\dot{r}/r$ is smaller than
$\Gamma \equiv  \dot{N}/N$.
Only for the ``stationary'' solution (\ref{85}) the ``physical'' particle
production rate coincides with $\Gamma $.
For $r={\rm const}$ the ``particle production'' is entirely connected with
the Q component.

In our basic setting the cosmic substratum is entirely made of
non-relativistic particles which are governed by an equilibrium distribution  
function.
The splitting (\ref{73})  has revealed that this ``generalized'' equilibrium  
description may be regarded as a two-fluid model of ``conventional'' and
``exotic'' matter.
The point is that also the ``exotic'' matter is described in terms of
``conventional'' matter particles, only that the property of being ``exotic''  
requires a ``production'' process.
If generated at a certain rate, ``conventional'' particles in the expanding  
universe effectively exhibit vacuum-like properties, which manifest
themselves macroscopically through a negative pressure.

Our fluid approach formally implies that a particle number is
attributed to the ``vacuum'' component as well.
At first sight the concept of a particle number of the vacuum might appear
obscure. In a fluid picture, however, it appears quite naturally
[cf. Ref. \cite{GaLeDeCQG}], as long as the interpretation of the first
moment (\ref{14}) of the distribution function as ``particle number flow'' is  
maintained.
As in any two-fluid picture,
the production rate $\Gamma $ naturally splits into
\begin{equation}
n \Gamma = n _{M}\Gamma _{M} + n _{Q}\Gamma _{Q}\ ,
\label{90}
\end{equation}
implying a split of the total particle number density
$n$ of the one-component model into $n = n _{M}+ n _{Q}$, by which the  
notions of a particle number density $n _{Q}$ of $Q$ ``particles'' and a  
corresponding change rate
$\Gamma _{Q}$ are introduced. $\Gamma _{M}=\dot{r}/r$ is the rate by which  
the CDM particle number changes, $n _{M}$   is the CDM particle number  
density.
A non-vanishing
$\Gamma $ does not necessarily imply a non-vanishing $\Gamma _{M}$ but may be   
a  feature of the fluid picture of ``vacuum matter'', characterized by
$\Gamma _{Q}\neq 0$.

\subsection{The case $B \propto -H ^{-1}$ }

In this subsection we focus on the previously discussed case
$B \propto - H^{-1}$ [cf. Eqs. (\ref{66})-(\ref{71})] for which the energy
density is given by
\begin{equation}
\rho
= \frac{\rho _{0}}{\mu +1}\left[\mu \left(1+z \right)^{3} + 1 \right]\ .
\label{91}
\end{equation}
Performing the splitting (\ref{73}), it is convenient to
replace
$r$ by
\begin{equation}
r \left(z \right) = \frac{\mu }{\mu + 1}f \left(z \right) \ ,
\mbox{\ \ \ }
r _{0} = \frac{\mu }{\mu + 1}\ ,
\mbox{\ \ \ }
f_{0} = 1\ .
\label{92}
\end{equation}
The energy density (\ref{91}) then decomposes into
\begin{equation}
\rho _{M} = \frac{\mu  }{\mu + 1}\rho _{0}\left(1+z \right)^{3}
f \left(z \right)\ ,
\mbox{\ \ \ }\mbox{\ \ \ }\mbox{\ \ \ }
\rho _{Q} = \frac{\rho _{0}}{\mu  + 1}
\left[1 - \left(f(z) - 1 \right) \mu \left(1+z \right)^{3}\right]\ .
\label{93}
\end{equation}
For the special case $f=1$ this splitting characterizes a non-interacting
mixture of
non-relativistic matter
and ``vacuum''.
This is just the $\Lambda $CDM model as already mentioned below
Eq. (\ref{69}).
In the general case $f = f \left(z \right)$ both components interact.
More specifically, there will be a decay of the Q component into matter.
The constant $\mu  $ is  the  ratio
\begin{equation}
\mu  = \frac{\rho _{M}\left(z=0\right)}{\rho _{Q}\left(z=0 \right)}\ .
\label{94}
\end{equation}
Here it is expedient to emphasize that the basic one-component dynamics
(\ref{67}) and (\ref{70}) is compatible with any continuous positive-definite
but otherwise arbitrary function $f$. The splitting into two components does 
not affect the quantity $\Pi/\rho$ at all.
However, different splittings may produce different perturbation spectra. In  
particular, isocurvature perturbations may occur which should be sensitive  
to the type of splitting.
Corresponding effects are expected to leave an imprint on the CMB anisotropies 
and are potential tools to discriminate between different choices.
Furthermore, the underlying two-component
dynamics is relevant for the coincidence problem.
The corresponding stationary solution $\rho_M/\rho_Q = {\rm const}$ is  
easily found. The condition
(\ref{85}), together with Eqs. (\ref{92}) and (\ref{71}), provides
\begin{equation}
f _{s}
= \frac{1}{\mu +1}\left[\mu + \left(1+z \right)^{-3} \right]\ ,
\label{95}
\end{equation}
where the subscript s again indicates stationarity according to Eq. (\ref{86}).
The obtained solution (\ref{95}) for $f$  is expected to describe the splitting
(\ref{93}) for large cosmological times.
For the corresponding asymptotic behavior of $\rho _{M}$ and $\rho _{Q}$ we
obtain
\begin{equation}
\left(\rho _{M} \right)_{s} = \frac{\mu \rho _{0}}{\left(\mu + 1 \right)^{2}}
\left[1 + \mu \left(1+z \right) ^{3}\right]
\label{96}
\end{equation}
and
\begin{equation}
\left(\rho _{Q} \right)_{s} = \frac{\rho _{0}}{\left(\mu + 1 \right)^{2}}
\left[1 + \mu \left(1+z \right)^{3} \right]\ ,
\label{97}
\end{equation}
respectively. These asymptotic expressions satisfy
\begin{equation}
\left(\frac{\rho _{M}}{\rho _{Q}} \right)_{s}
= \mu = \frac{r _{0}}{1 - r _{0}} = {\rm const}\ .
\label{98}
\end{equation}
Both components red-shift at the same rate. There is a permanent energy
transfer from the Q component to the matter component. Without transfer
$\rho _{M}$ would red-shift as $a ^{-3}$  while $\rho _{Q}$ would remain
constant. The transfer makes $\rho _{Q}$ red-shift also and, on the other
hand, $\rho _{M}$ to red-shift at a lower rate than without transfer.
In general, the red-shifts differ. For a specific amount of transfer,
however, given by the expression (\ref{95}) for $f$, the rates just coincide.
We conclude that a fixed ratio $\rho _{M}/\rho _{Q}$ is compatible with cosmic
anti-friction.
If our universe follows such an asymptotic solution presently,
this should manifest itself by a violation of the CDM
particle number. If we are still far from the asymptotic regime, the CDM
particle production might be negligible.
These considerations of the coincidence problem give rise to a picture
according to which an initial cosmological term did not completely vanish
during the early stages of the cosmological evolution but still exists and
even continues to decay.
At large cosmological times not only the cosmological term itself becomes
dynamically relevant again but also its decay properties may be essential for  
the asymptotic state of the universe (ignoring here a recently discussed
scenario with a less bleak eschatological picture \cite{JBarr}).

\section{Summary}

We have introduced the concept of cosmic anti-friction to discuss the possible
origin of an accelerated expansion of the present universe.
Cosmic anti-friction relies on a one-component picture of the cosmic  
substratum which is
regarded as a self-interacting gas of non-relativistic particles in
(generalized) equilibrium.
Together with simple  assumptions about the interaction rate it allowed us  
to establish exactly solvable models of the cosmological dynamics.
Cosmic anti-friction leads to a
negative bulk pressure  which may well account for the magnitude-redshift
data of type Ia supernovae. The $\Lambda $CDM scenario is recovered as
a special case of cosmic anti-friction.
For the models of subsections IV.B. and IV.C., which are consistent both  
with the SN Ia observations and with CMB anisotropy data, we find the  
beginning of the phase of accelerated expansion at redshifts $z _{{\rm  
acc}}\approx 1.8$ and
$z _{{\rm acc}}\approx 0.7$, respectively.
The one-component dynamics may be split into a two-fluid mixture
in which a Q matter component decays into CDM.
For a suitable decay rate there exists an attractor solution characterized
by a fixed ratio of the energy densities of both components, which indicates  
a possible solution of the coincidence problem.
As for scenarios with a cosmological constant or quintessence, the  
microphysical evidence for our models remains open.
This shortcoming seems presently unavoidable and reflects our basic  
ignorance concerning the substance our Universe is made of.

\acknowledgments
This paper was supported by the Deutsche Forschungsgemeinschaft,
the Spanish Ministry of Science and Technology (Grant
BFM 2000-0351-C03-01) and
NATO. D.J.S. thanks the Austrian Academy of Sciences
for financial support.

\end{document}